\documentclass[final,1p,times]{elsarticle} 
\usepackage{graphicx} 
\usepackage{epstopdf}
\usepackage{amssymb} 
\usepackage{amsthm} 
\usepackage{lineno} 

\journal{Nuclear Physics A} 
\begin{document} 
\setlength{\textheight}{9.25in}
\setlength{\topmargin}{-0.25in}
\begin{frontmatter} 


\title{Centrality Dependence of Two-Particle Correlations in Heavy Ion Collisions}

\author{George S.F.~Stephans$^5$ for the PHOBOS collaboration\\
B.Alver$^5$,
B.B.Back$^1$,
M.D.Baker$^2$,
M.Ballintijn$^5$,
D.S.Barton$^2$,
S.Basilev$^5$,
B.D. Bates$^5$,
R.Baum$^8$,
B.Becker$^2$
% ...
}

\address{
$^1$~~Argonne National Laboratory, Argonne, IL 60439, USA\\
$^2$~~Brookhaven National Laboratory, Upton, NY 11973, USA\\
$^3$~~Institute of Nuclear Physics, Krak\'{o}w, Poland\\
$^4$~~Jagellonian University, Krak\'{o}w, Poland\\ 
$^5$~~Massachusetts Institute of Technology, Cambridge, MA 02139, USA\\
$^6$~~National Central University, Chung-Li, Taiwan\\
$^7$~~University of Illinois at Chicago, Chicago, IL 60607, USA\\
$^8$~~University of Maryland, College Park, MD 20742, USA\\
$^9$~~University of Rochester, Rochester, NY 14627, USA\\
\vspace{-30pt}
}

\begin{abstract} 
Data from the PHOBOS detector have been used to study two-particle correlations over a broad range of pseudorapidity. A simple cluster model parameterization has been applied to inclusive two-particle correlations over a range of centrality for both Cu+Cu and Au+Au collisions at $\sqrt{s_{_{NN}}}=200$~GeV. Analysis of the data for Au+Au has recently been extended to more peripheral collisions showing that the previously-observed rise in cluster size with decreasing system size eventually reaches a maximum value. Model studies have been used to quantify the significant effect of limited detector acceptance on the extracted cluster parameters. In the case of Au+Au, correlations between a trigger particle with $p_T>2.5$~GeV and inclusive associated particles have also been studied. These reveal the presence of a `ridge' at small relative azimuthal angle which extends with roughly constant amplitude out to the largest relative pseudorapidity studied. The large phase-space coverage of the PHOBOS detector has enabled a quantitative understanding of the so-called `ZYAM' parameter used in the subtraction of the contribution of elliptic flow to these triggered correlations.
\end{abstract} 

\end{frontmatter} 

\setlength{\textheight}{9.25in}
\setlength{\topmargin}{-0.25in}



Studies of single-particle distributions extracted using data from ultra-relativistic heavy ion collisions have provided a wealth of information concerning the global properties of these interactions. Two-particle correlations can provide more detailed information, in particular probing the characteristics of the particle production process itself. The PHOBOS detector at the Relativistic Heavy Ion Collider (RHIC) has been used to study correlations in which both particles in a pair are chosen from the inclusive distribution \cite{Wei1, Wei2} as well as those in which inclusive particles are correlated with a trigger particle selected to have $p_T>2.5$~GeV \cite{TrigCorr}. 

\begin{center}
\begin{figure}[ht]
\includegraphics[width=0.48\textwidth]{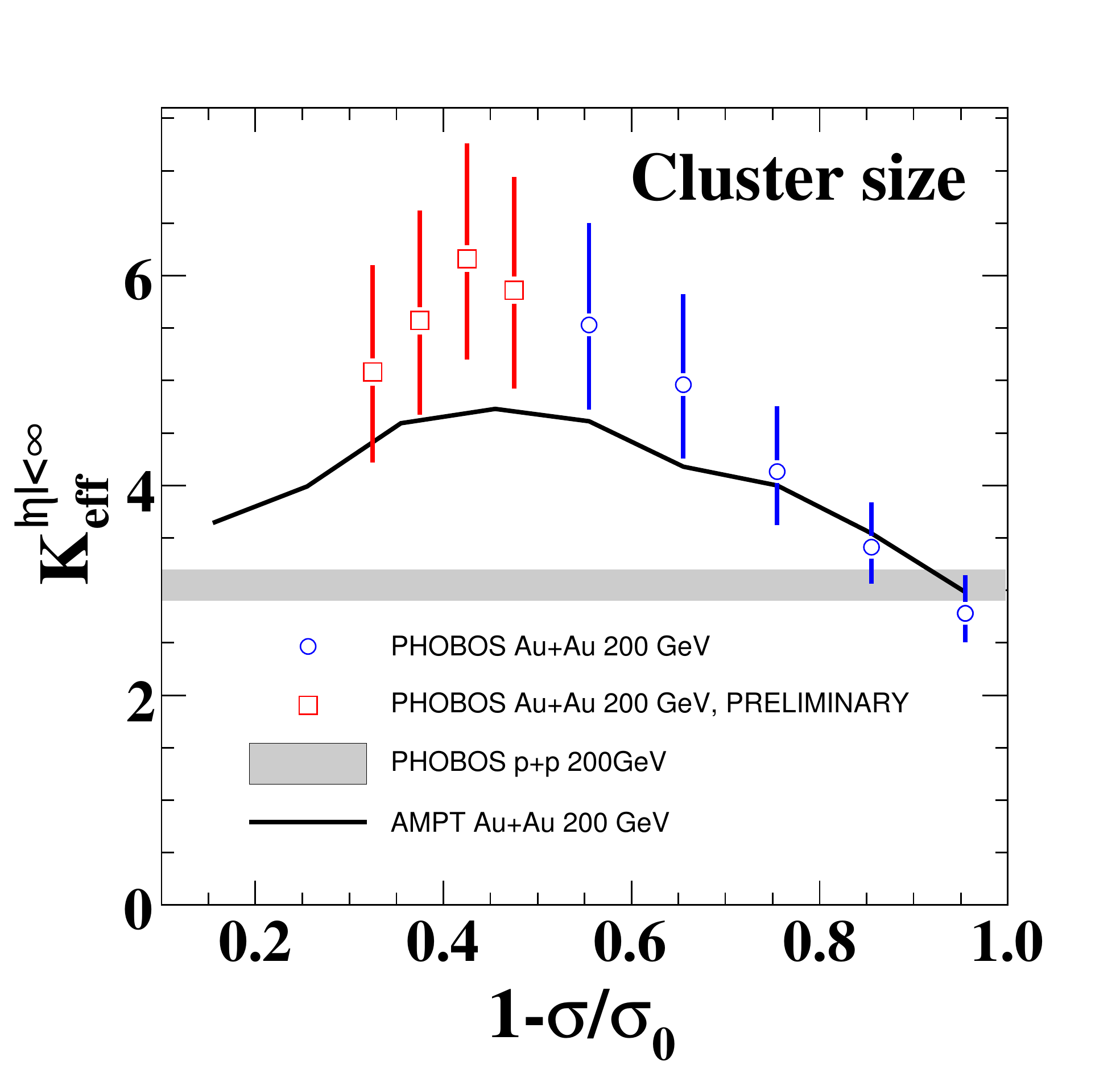}
\hspace{0.04\textwidth}
\includegraphics[width=0.46\textwidth]{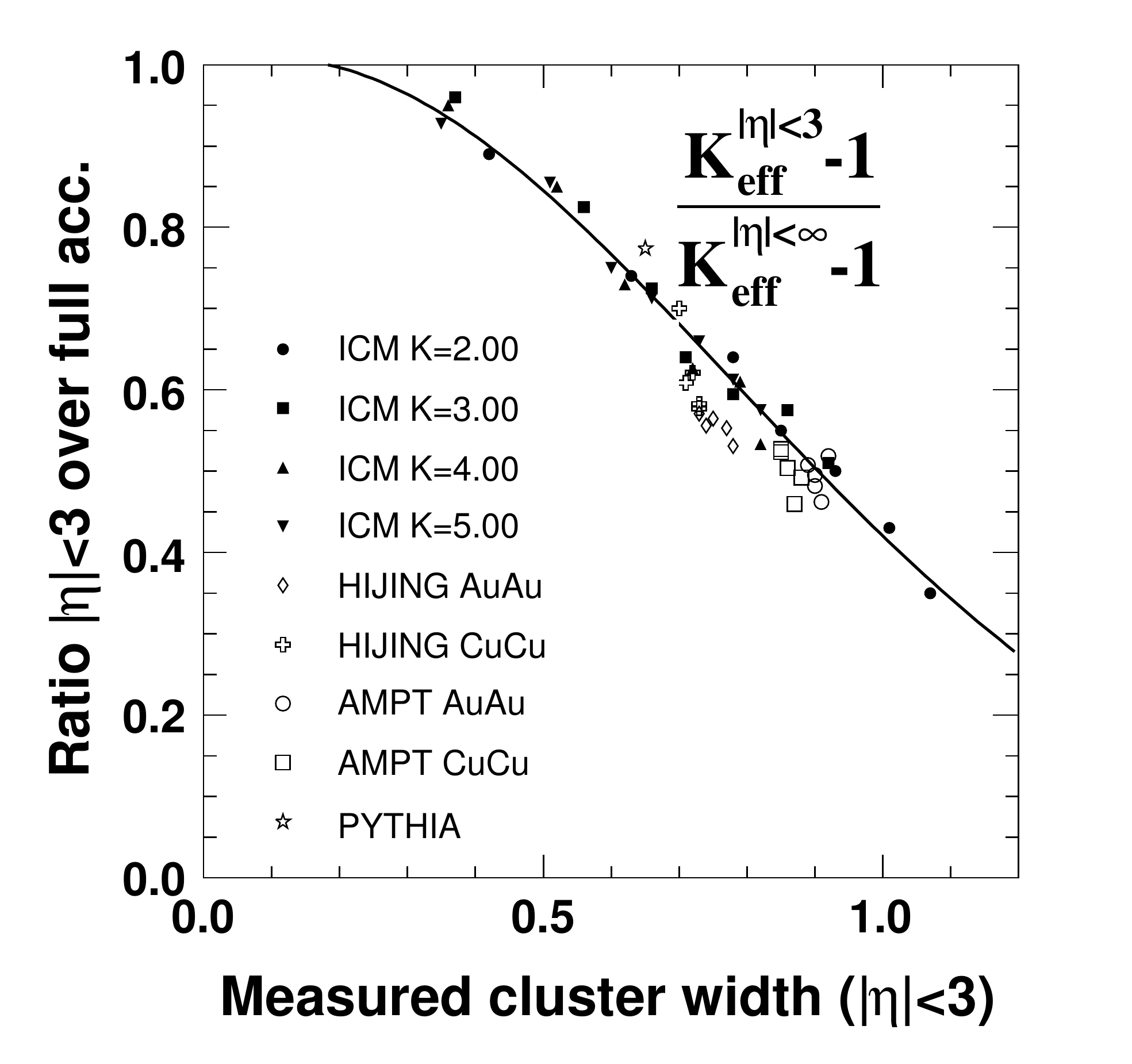}
\caption{(color online) Left panel: Extracted cluster sizes, corrected for acceptance effects, for Au+Au collisions at $\sqrt{s_{_{NN}}}=200$~GeV as a function of the fraction of the total inelastic cross section (equal to 1.0 for the most central collisions). Error bars include both statistical and systematic effects. The gray band shows the acceptance-corrected cluster size for p+p collisions. The black line is the cluster size predicted by the AMPT model. Right panel: The ratio of the cluster size extracted using only the particles within a limited detector acceptance over the size found using all particles is plotted for a variety of models as a function of the cluster width within the limited acceptance. Note that heavy ion collisions at $\sqrt{s_{_{NN}}}=200$~GeV typically have measured cluster widths of about 0.7$-$1.0. See text for discussion.}
\label{2-part}
\end{figure}
\end{center}

Inclusive correlations were analyzed using a simple cluster model parameterization for particles emitted in collisions of Au+Au and Cu+Cu at $\sqrt{s_{_{NN}}}=200$~GeV \cite{Wei2}. The average number of emitted particles (cluster size, $K_{\rm eff}$) and their spread in pseudorapidity (cluster width, $\delta$) were extracted for interactions over a range of centrality. It was found that the cluster sizes are quite large and increase with decreasing system size. Comparing results for Au+Au and Cu+Cu, similar cluster sizes are observed for collisions at the same fraction of the total inelastic cross section, as opposed to systems with the same number of participants. The left panel of Fig.~\ref{2-part} shows data for the cluster sizes versus centrality in Au+Au which have recently been extended to more peripheral collisions. The rise in cluster size is seen to saturate and possibly even begin to decrease in the most peripheral collisions. Continuation of this trend might result in values for the cluster size close to that seen for p+p collisions which is shown by the gray band in the figure.

Model calculations were analyzed to study the effect of limited detector acceptance on the extracted cluster parameters \cite{Wei2}. A range of very different models was studied, including both simple independent cluster production and more complicated dynamical calculations. Specifically, the analysis determined the ratio of the cluster size extracted using a limited set of particles over the size found using all particles. As shown in the right panel of Fig.~\ref{2-part}, this ratio was very similar in all of the models when comparing systems with similar cluster width (i.e. the RMS of the separation in pseudorapidity between pairs of particles from a cluster). This result seems intuitively reasonable since for broader particle emission from a particular cluster, it is more likely that one or more of the daughter particles will fall outside the detector acceptance. What was not expected was the relatively large magnitude of the effect even for the PHOBOS detector which has the largest pseudorapidity acceptance at RHIC. In order to extract the correction to apply to experimental data, ratios were determined as a function of the cluster width found using only the restricted set of particles. For the typical cluster widths seen in heavy ion collisions at $\sqrt{s_{_{NN}}}=200$~GeV, a detector extending over $|\eta|<3$ finds only about half of $K_{\rm eff}-1$ where $K_{\rm eff}$ is the cluster size. Similar, although smaller, correction factors were found for the cluster width. Data from a detector with a  pseudorapidity acceptance smaller than PHOBOS would have an even more limited sensitivity to the effect of this aspect of particle production.

Correlations between a trigger particle with $p_T>2.5$~GeV and inclusive associated particles were analyzed for collisions of Au+Au at $\sqrt{s_{_{NN}}}=200$~GeV over a range of centralities \cite{TrigCorr}. A particularly interesting feature of preliminary triggered correlations data from other experiments was the so-called `ridge', an enhancement at small relative azimuthal angle which was extended in relative pseudorapidity compared to the same correlations from p+p interactions \cite{STARridge}. 
\begin{figure}[ht]
\includegraphics[width=0.53\textwidth]{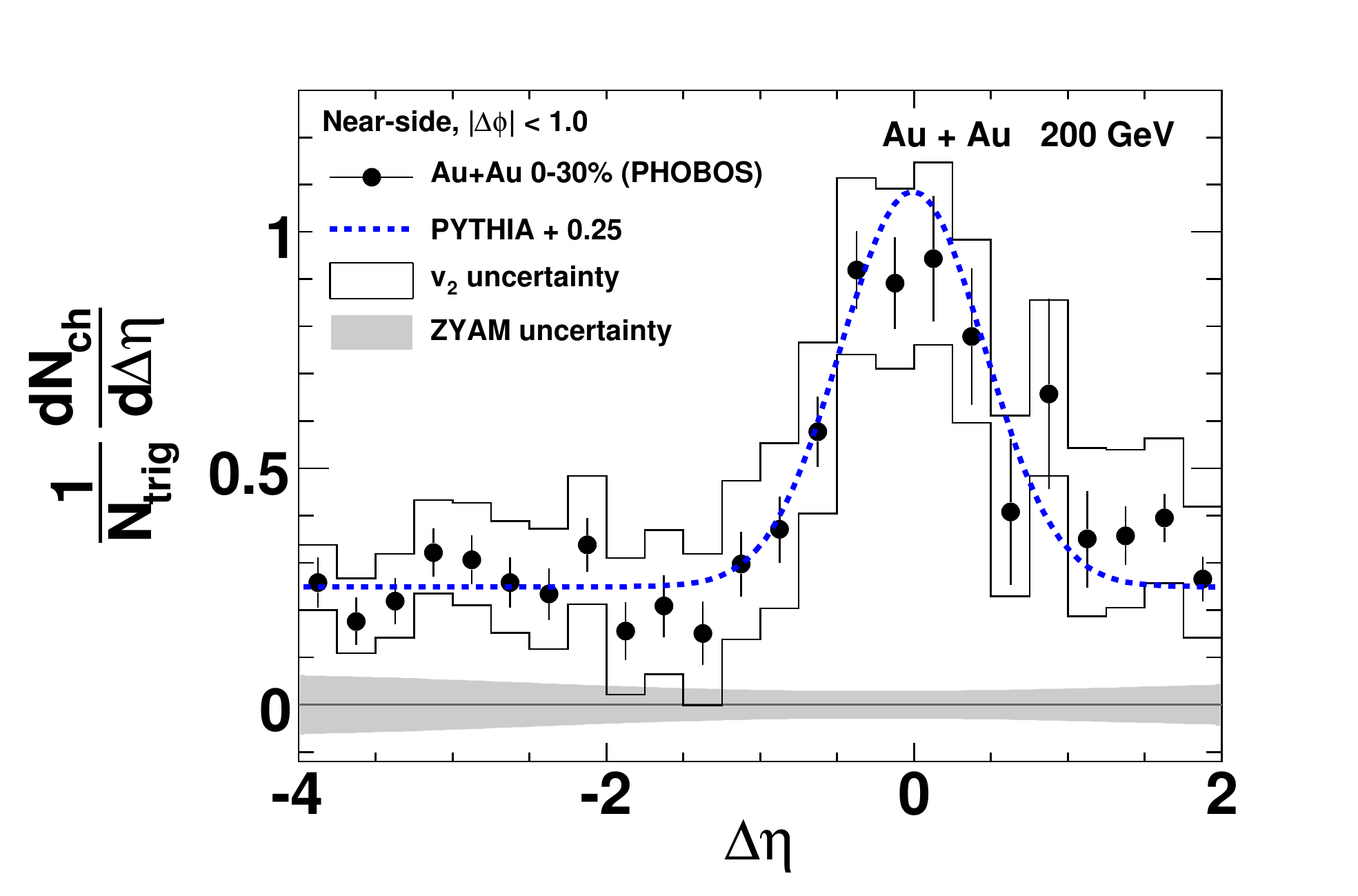}
\includegraphics[width=0.45\textwidth]{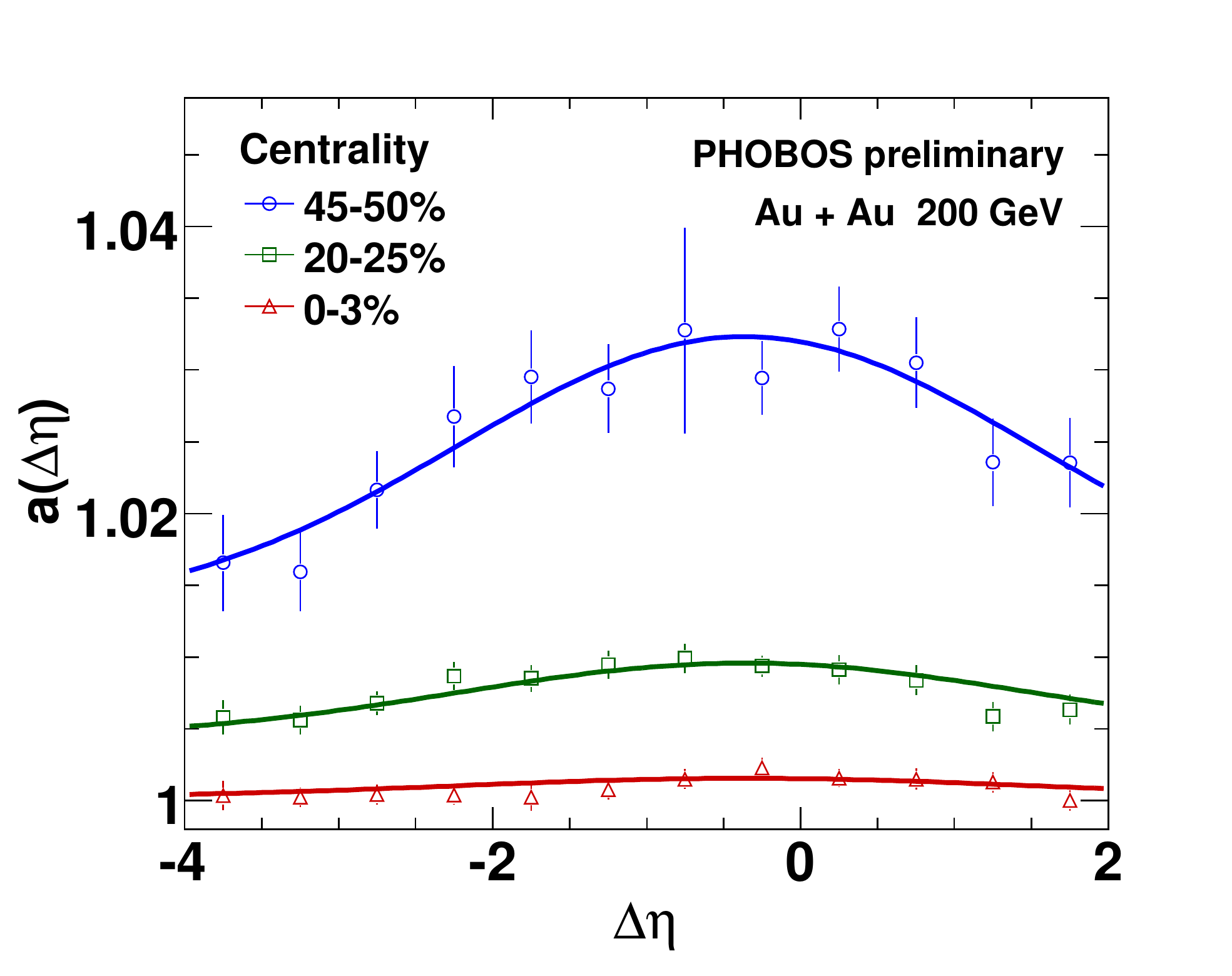}
\caption{(color online) Left panel: Correlated particle yield for trigger particles with $p_T>2.5$~GeV emitted in central Au+Au collisions at $\sqrt{s_{_{NN}}}=200$~GeV as a function of relative pseudorapidity. Error bars are statistical errors and boxes represent the systematic error due to the uncertainty in the elliptic flow subtraction. The gray band shows the systematic error due to the normalization between background and signal distributions. The dashed line shows the PYTHIA prediction for p+p at $\sqrt{s}=200$~GeV shifted up by a constant offset. Right panel: The normalization between signal and background distributions used in the elliptic flow subtraction for Au+Au collisions at $\sqrt{s_{_{NN}}}=200$~GeV in three centrality ranges is plotted as a function of relative pseudorapidity. See text for discussion.}
\label{trig-corr}
\end{figure}
The unique feature of this analysis using PHOBOS data is the very broad accessible range in relative pseudorapidity between the two particles. Among other results, it was discovered that this `ridge' extends to the largest pseudorapidities analyzed. As shown in the left panel of Fig.~\ref{trig-corr}, the data are consistent with a constant height of the ridge over the entire range. The dashed line in that panel shows the PYTHIA prediction for p+p at the same center-of-mass energy shifted up by a constant value of 0.25 associated particles per unit of relative pseudorapidity.

In extracting the triggered correlation functions, it is necessary to subtract the effect of elliptic flow as shown in Eq.~\ref{CorrYld} where the left side is the correlated particle density. The factor $B(\Delta\eta)$ is a normalization based on $dN/d\eta$; $s(\Delta\phi,\Delta\eta)$ and $b(\Delta\phi,\Delta\eta)$ are the signal (pairs from a single event) and background (pairs from mixed events) distributions, respectively, both normalized to the number of trigger events; and $V(\Delta\eta)$ is a convolution of the elliptic flow magnitude for the trigger and associated particles (see Ref.~\cite{TrigCorr} for details).
\begin{equation}
\frac{1}{N_{trig}} \frac{d^{2}N_{ch}}{(d\Delta\phi) (d\Delta\eta)} = B(\Delta\eta)  \cdot \left[\frac{s(\Delta\phi,\Delta\eta)}{b(\Delta\phi,\Delta\eta)}  - a(\Delta\eta) [ 1+2V(\Delta\eta) cos(2\Delta\phi) ]\right].
\label{CorrYld}
\end{equation}

In the absence of effects which distort the relative normalization of the signal and background distributions, the factor $a(\Delta\eta)$ would not be needed. The right panel of Fig.~\ref{trig-corr} shows the value of $a$ as a function of relative pseudorapidity for Au+Au collisions in several ranges of centrality. The data points were found by matching the uncorrected correlation function and the predicted flow signal using a variety of techniques \cite{TrigCorr}, one of which was the so-called ZYAM or zero yield at minimum \cite{ZYAM}. In previous analyses, this adjustment was treated as a somewhat arbitrary value. However, the uniquely broad range of the PHOBOS detector allowed a quantitative understanding of the origin of this factor. Three features are clear in the right panel of Fig.~\ref{trig-corr}, namely that $a$ is typically very close to unity, that the deviation from one decreases for more central collisions, and that the dependence on relative pseudorapidity can be characterized as a Gaussian peak on top of a constant value. The lines in the figure are the result of a fit to the full set of data points (i.e. all centralities and relative pseudorapidities) using a parameterization including both constant and Gaussian terms. 

The component of the $a$ factor which is independent of pseudorapidity is due to a simple trigger bias; events at the higher end of a centrality bin are more likely to contain the required trigger particle. As a result, the distribution of signal events within a given centrality bin is biased towards the higher multiplicity end, as compared to the background events. By simple combinatorics, this shift results in an artificial enhancement in the number of pairs at all relative pseudorapidities. Both the magnitude and centrality dependence of this pseudorapidity-independent component of $a$ can be reproduced quantitatively based on the widths and average particle multiplicities of the centrality bins used in this analysis. 

The Gaussian component is simply a reflection of the cluster-like particle-production processes discussed above, which impart a relative-pseudorapidity-dependent normalization between the signal and background distributions. Both the Gaussian width and the centrality dependence of the Gaussian peak height found in fitting the $a$ values are consistent with those found in the inclusive two-particle analysis. Note that the definition of the $\delta$ parameter in Ref.~\cite{Wei2} differs from the width of the Gaussian in $\Delta\eta$ found when fitting $a$ by a factor of $\sqrt{2}$. Comparing inclusive and triggered correlations, the magnitude of the underlying cluster effect is not necessarily the same since in one case there is a requirement of a high transverse momentum trigger particle. This difference was reflected in the fit by a single overall  normalization of the Gaussian peak height. The centrality dependence of this peak amplitude includes the trivial multiplicity-dependent dilution of correlated versus uncorrelated pairs. Note that the correlation functions used in the cluster analysis differ from those used in the triggered correlation study in that this multiplicity-dependent dilution is explicitly removed in the cluster analysis (see Eq.~1 of Ref.~\cite{Wei2}). However, there is also the non-trivial centrality dependence of the cluster size itself as seen in the left panel of Fig.~\ref{2-part} and described in detail in Ref.~\cite{Wei2}.

These expectations for the characteristics of the $a$ factor were not directly included in the fit although they did influence the choice of functions. As one example, the Gaussian amplitude included an explicit (1/multiplicity) dependence. Characteristics of the data, including a varying centrality bin width, precluded the use of a simple analytic form, although predictions could be easily calculated bin by bin. The final results of the fits were compared to the predictions based on the effects discussed above and found to be in good quantitative agreement. As is clear in the right panel of Fig.~\ref{trig-corr} and as results from, among other effects, the dramatic impact of detector acceptance shown in the right panel of Fig.~\ref{2-part}, this quantitative description of the various sources of offset between background and signal distributions in triggered correlations would not have been possible without the broad pseudorapidity coverage of the PHOBOS detector.

In summary, PHOBOS data have been used to significantly enhance our knowledge of the nature of particle correlations in heavy ion collisions at the top RHIC energy. The data imply that particles are emitted from surprisingly large clusters with a non-trivial centrality dependence which has been recently extended towards smaller systems. The significant effect of detector acceptance, even in a detector as broad as PHOBOS, has been quantified. The `ridge' at small relative azimuthal angle in triggered correlations has been found to extend over the full relative pseudorapidity range studied. Finally, the non-trivial normalization required when subtracting the effect of elliptic flow on two-particle correlations has been understood quantitatively.



\end{document}